# Quantum Enhanced Josephson Junction Field-Effect Transistors for Logic Applications

W. Pan[1], A.J. Muhowski[2], W.M. Martinez[2], C.L.H. Sovinec[2], J.P. Mendez[2], D. Mamaluy[2], W. Yu[2], X. Shi[2], K. Sapkota[2], S.D. Hawkins[2], and J.F. Klem[2]

[1] Sandia National Laboratories, Livermore, CA 94550, USA

[2] Sandia National Laboratories, Albuquerque, NM 87123, USA

Abstract:

Josephson junction field-effect transistors (JJFETs) have recently re-emerged as promising candidates for superconducting computing. For JJFETs to perform Boolean logic operations, the so-called gain factor $\alpha_R$ must be larger than 1. In a conventional JJFET made with a classical channel material, due to a gradual dependence of superconducting critical current on the gate bias, $\alpha_R$ is much smaller than 1. In this Letter, we propose a new device structure of quantum enhanced JJFETs in a zero-energy-gap InAs/GaSb heterostructure. We demonstrate that, due to an excitonic insulator quantum phase transition in this zero-gap heterostructure, the superconducting critical current displays a sharp transition as a function of gate bias, and the deduced gain factor $\alpha_R \sim 0.06$ is more than 50 times that ($\sim 0.001$) reported in a classical JJFET. Further optimization may allow achieving a gain factor larger than 1 for logic applications.

In order to meet the enormous power-consumption demand for computation, information, and communication, the microelectronics industry has embarked on a long and extremely fruitful journey in aggressively shrinking the size of logic and memory transistors. As shown in a past study [1], a power reduction greater than 55% was achieved when moving from the 28 nm high-k transistor node to the 16 nm Fin-FET (field-effect transistor) transistor node at that time. However, the scaling is fast approaching atomic-scale, materials-derived limits, which will present tremendous challenges in transistor operation. Thus, exploration of novel materials and device architectures to meet future power demand is warranted.

Superconducting computing is promising for low-energy, power efficient circuit applications. It has been estimated that its power consumption could be at least 10 times lower than the most efficient exascale conventional silicon computing for the same performance [2]. In addition, using superconducting materials as interconnects can further reduce power consumption. Since the early 1990s, rapid single flux quantum (RSFQ) circuits have been a mainstay in superconducting computing [3]. However, known issues, such as a large static power dissipation, have prevented RSFQ circuits from being widely deployed in general computing applications. Moreover, RSFQ circuits cannot take advantage of vast digital logic design software currently available.

Josephson junction field-effect transistors (JJFETs) have recently re-emerged as promising candidates for superconducting computing [4-6]. JJFETs are particularly useful for low power applications as, in the superconducting regime, they are operated with zero-voltage drop across their source and drain electrodes. The feasibility of using JJFETs for Boolean logic operations has been explored [5], and a high noise margin for logic inputs has been demonstrated [6]. Additionally, unlike RSFQ, JJFET circuits are compatible with existing logic design software for conventional silicon CMOS.

A JJFET (Fig.1a) is very similar to a Si-MOSFET, except that the source and drain electrodes are replaced by a superconducting material. In this Josephson junction structure, a supercurrent $I_c$ can exist in the channel due to the overlap between the Cooper pair wave-functions in the source and drain electrodes. In general, $I_c \propto \exp(-L/\xi_c)$ [7], where L is the channel length and $\xi_c$ the carrier coherence length. Usually, $\xi_c$ displays a power law dependence on carrier density (n), $\xi_c \propto n^\gamma$. Applying a gate bias $V_g$ in a JJFET changes n. This, in turn, changes $I_c$. At high gate bias voltages, or $V_g > V_t$ (where $V_t$ is the threshold voltage), both n and $\xi_c$ are nonzero. A supercurrent $I_c$ can appear in the channel (i.e., the device is in the superconducting state). Consequently, the voltage-drop between the source and drain $V_{dc} = 0$; this can be viewed as the "0" state. When $V_g$ approaches $V_t$, n, $\xi_c$, and $I_c$ all approach zero. There is no supercurrent in the channel and the device is in a resistive state. Under a finite bias current $I_{bias}$, $V_{dc}$ is non-zero; this can be viewed as the "1" state.

For JJFETs to perform Boolean logic operations, the so-called gain factor $\alpha_R = dI_c/dV_g \times \pi\Delta/I_c$ must be larger than 1 [5]. Here $\Delta$ is the superconducting gap. In conventional JJFETs made with a classical semiconductor channel material, $\gamma$ takes a value of 0.5 in the clean limit [7], or $\xi_c \propto n^{0.5}$. Since $n \propto V_g-V_t$ in a typical FET device, $\xi_c \propto (V_g-V_t)^{0.5}$ (as shown in Fig. 1b). As a result, $I_c$ demonstrates a gradual dependence on $V_g-V_t$ (Fig. 1c), yielding a small $dI_c/dV_g$ value. In order to achieve $\alpha_R > 1$ with such a small $dI_c/dV_g$, a superconducting transition temperature ($T_c$) as high as, for example, 400K is needed, far exceeding critical temperatures in any superconducting materials. As such, it is currently impossible to use conventional JJFETs for logic operations. In fact, this contributes to JJFETs remaining a laboratory research topic despite being proposed more than 40 years ago [4].

Here, we propose a new scheme that can enable realization of a large gain factor $\alpha_R > 1$, thus making JJFETs a prime candidate for a lab-to-fab transition. In order to achieve $\alpha_R > 1$ with practically realizable moderate $T_c$ superconductors, we need significantly larger $\gamma$ in the channel material, for example $\xi_c \propto (V_g-V_t)^5$ (Fig. 1b). As presented in Fig. 1c, such large $\gamma$ can induce a sharp increase in $I_c$ with $V_g-V_t$, thus a large value of $dI_c/dV_g$. A large value of $dI_c/dV_g$ then allows to implement a superconductor with a moderate $T_c$ for achieving $\alpha_R > 1$. To realize large $\gamma$, we propose to replace the classical channel materials (such as conventional semiconductor) with a novel materials stack that can undergo a sharp quantum phase transition via tuning $V_g$. Due to the nature of collective phenomena in a quantum phase transition, $\xi_c$ can display a strong n or $V_g$-$V_t$ dependence. In the following, we will demonstrate that the excitonic insulator (EI) phase transition, which was recently reported in type-II InAs/GaSb heterostructures [8-11], can be utilized to enable this kind of quantum enhanced [12] JJFETs.

The concept of the EI transition was proposed more than 60 years ago [13]. In a semimetal, electrons and holes form pairs (or excitons) bound by Coulomb attraction, and the coherent condensation of these excitons then leads to a new ground state. Later, it was predicted [14] that in a narrow-gap semiconductor if the binding energy of excitons ($E_B$) exceeds the energy gap ($E_g$), excitons form a Bose-Einstein condensate below a critical temperature. Consequently, a larger gap is opened, leading to a new insulating ground state, now dubbed the EI phase. Recently, we reported the observation of an EI phase in a zero-gap InAs/GaSb heterostructure [8]. In this heterostructure, the lowest electric subband ($E_0$) in the conduction band and the highest hole subband in the valence band ($H_0$) almost perfectly touch each other, resulting in an energy gap $E_g \approx 0$. Usually, a semi-metallic state is expected if $E_g = 0$. Surprisingly, at low temperature (T) in this zero-gap InAs/GaSb heterostructure, we observe a huge resistance peak, ~ 600 k$\Omega$, when $V_g$ sweeps into the charge neutrality region [8], a regime in which the densities of electrons and holes are roughly equal. This result from Ref. [8] and our further studies demonstrate that this huge resistance is due to the formation of an EI phase. Indeed, with $E_g \approx 0$ the condition $E_g < E_B$ is readily realized, and an EI is expected. Importantly, this EI quantum phase transition, as we will demonstrate below, can enable a strong $V_g$-$V_t$ dependence for $\xi_c$ (large $\alpha_R$), thus making the EI transition enhanced JJFETs suitable for logic applications.

Below, we present results in a JJFET made of this zero-gap InAs/GaSb heterostructure with tantalum (Ta) as the source and drain electrodes. We show that the superconducting critical current in the JJFET is zero when $V_g$-$V_t \leq 0.23$V, but sharply jumps to a finite value at $V_g$-$V_t$ = 0.24V and then increases slowly as $V_g$-$V_t$ continues to increase. The gain factor is calculated to be ~ 0.06. Though still less than 1, it is more than 50 times that recently reported in a conventional JJFET made of InAs quantum wells [15]. With further optimization, a sharper excitonic insulator transition can be achieved, and a larger gain factor can be expected.

The growth structure of the zero-gap InAs/GaSb heterostructure is the same as sample B in Ref. [8]. The thicknesses of InAs and GaSb quantum wells are 10 and 5 nm, respectively. At these thicknesses, the energy-gap between $E_0$ and $H_0$ is nearly zero [8]. To fabricate JJFETs, we first utilize photolithography and wet etching methods to realize a mesa (green colored in Fig. 2a) with a size of 100 μm × 4 μm. E-beam lithography is then used to define superconducting Ta electrodes and wet etched to produce trenches before the ~ 120 nm thick Ta electrodes (marked in Fig. 2a) are sputter-deposited. The channel length of the Josephson junction measured in this work is about 500 nm. Finally, an atomic-layer-deposited $Al_2O_3$ thin film (80 nm thick) and a Ti/Au (10 nm/300 nm) gate metal stack are followed to complete the FET device fabrication. DC measurements are carried out using a Keithley 238 source meter. For low frequency (~ 11Hz) phase lock-in measurements, Stanford Research Systems SR830's are employed. All of the measurements are taken at T = 11 mK except for the T dependence in Fig. 2c.

Fig. 2a shows an optical image of the device. In Fig. 2b, the sample conductance σ as a function of $V_g$-$V_t$ is displayed in a log-log plot. The charge neutrality point occurs at $V_g$=-0.7V, and we take this voltage as the threshold voltage $V_t$. Three regimes are identified. For $V_g$-$V_t$ < 0.2V, σ ~ 0.5 mS. In the high $V_g$-$V_t$ (> 0.4V) regime, the device is in the classical regime and σ = $ne^2τ/m^*$ ∝ $(V_g-V_t)^1$ ∝ $n^1$ (or τ ∝ $n^0$). Here, e is the electron charge, τ the transport time, and $m^*$ effective mass. Assuming the coherence length is roughly equal to the mean free path ($l_{mfp}$ = $V_F$×τ = $ℏk_Fτ/m^*$), we obtain $ξ_c$ ∝ $n^{0.5}$. Here, $V_F$ is the Fermi velocity, $k_F$ ∝ $n^{0.5}$ the Fermi wave factor for two-dimensional gas. Between 0.2 and 0.4V in the EI transition regime, σ ∝ $(V_g-V_t)^3$ ∝ $n^3$. Assuming σ = $ne^2τ/m^*$ still holds, then τ ∝ $n^2$. This translates to $ξ_c$ ∝ $n^{2.5}$, much stronger than that in the classical regime. We emphasize here that it is this strong density dependence that greatly enhance the gain factor in this JJFET.

Fig. 2c shows the four-terminal sample resistance $R_{xx}$ as a function of T. $R_{xx}$ is nearly constant (~ 1100 Ω) at high temperatures. It drops quickly to ~ 600 Ω at T ~ 1.25 K when Ta becomes superconducting. From this transition temperature, we deduce a superconducting gap $\Delta = 1.75k_BT_c = 0.19$ meV. $R_{xx}$ continues to decrease with decreasing temperature and saturates at a value of ~ 250 Ω when T < 0.8 K.

In Figure 2d, we show the DC current-voltage (I-V) characteristics taken at $V_g = 0$ V. In the regime of large DC current bias $I_{bias}$, the Josephson junction is in the normal state, and the voltage-drop between the source and drain $V_{dc}$ is linear with $I_{bias}$. The calculated differential resistance $dV_{dc}/dI_{bias}$ is ~ 580 Ω, consistent with the resistance value immediately after the sudden drop in Fig. 2b. Around the zero-bias current, the I-V curve again is linear but with a smaller slope. Correspondingly, the differential resistance is lower, ~ 230 Ω. This small resistance, or high conductance, is due to the superconducting proximity effect. We note that in this Josephson junction, a true supercurrent state (i.e., $V_{dc} = 0$) is not observed. This is probably due to a long separation, ~ 500 nm, between the two superconducting electrodes. Nevertheless, a "knee" like behavior between the low-bias developing supercurrent state and the high-bias normal state allows us to define a quasi-critical current based on the crossing point of two linear extrapolations as shown in Fig. 2d.

The main results of this work are displayed in Fig. 3. In Fig. 3a, we show the I-V curves at a few selected gate voltages in the EI transition regime. The knee-like behavior is clearly observable down to $V_g = -0.46$ V (or $V_g-V_t = 0.24$ V), though it becomes weaker as $V_g$ approaching $V_t$. Beyond -0.47 V (or $V_g-V_t = 0.23$ V), the I-V curve is linear, and no knee-like behavior is observed. In other words, the critical current is zero. In Fig. 3b, we determine $I_c$ using another method by examining $dV_{dc}/I_{dc}$. In the presence of a knee-like behavior, the derivative is expected to jump from a low value to a high value at the critical current. Indeed, this is observed for $V_g-V_t \geq 0.24$ V, from which we take the current value at the middle point in the jump (indicated by the up-arrows) as the $I_c$. For $V_g-V_t \leq 0.23$ V, no jump is observed. In other words, $I_c = 0$. In Fig. 3c, we plot $I_c$ obtained from both methods as a function of $V_g-V_t$. $I_c$ is zero for $V_g-V_t \leq 0.23$V. It jumps sharply to a non-zero value of ~ 0.33 µA from 0.23 to 0.24V, and then increases slowly as $V_g$-$V_t$ continues to increase. A linear fit in the sharp jump regime yields a value of $dI_c/dV_g = 33$ µA/V. Using this value together with the superconducting gap of $\Delta = 0.19$ meV, we deduce a gain factor of $\alpha_R \approx 0.06$ in this JJFET.

Though still less than 1, the gain factor obtained is over 50 times that (~ 0.001) recently obtained in a conventional JJFET made of InAs quantum wells [15]. Two potential improvements can be pursued for achieving a larger $\alpha_R$. First, the Josephson junction length could be reduced so that a true supercurrent state can be realized. Also, we can replace Ta with aluminum, which has been shown to form an excellent interface with InAs quantum wells [16]. Second, the heterostructure growth parameters can be optimized so that a sharper excitonic insulator transition and, thus, a larger $dI_c/dV_g$, will be achieved. With these improvements, we are optimistic that the EI quantum phase transition enhanced JJFETs will be primed for Boolean logic operations.

We add in passing that InAs/GaSb heterostructure is an ideal material system for achieving quantum enhanced JJFETs due to its high mobility, band-engineering tunability and compatibility with current microelectronics processing technologies. Moreover, its Fermi level pinning occurs above the InAs conduction band edge. Consequently, an electronically transparent interface can be easily achieved between a superconductor and InAs/GaSb. This is not the case in GaAs and Si, in which the known issue of a finite depletion length makes a transparent interface difficult [17]. In terms of band engineering, we note that it is possible to design unique InAs/GaSb heterostructures with flat conduction and valence bands [18]. It has been reported that the exciton condensate can be very sharp in flat-band bilayer systems [19]. Finally, we point out that the sharp excitonic insulator transition observed in other material systems, such as transition metal dichalcogenide semiconductor $WSe_2$/$MoSe_2$ double layers [20] and $WTe_2$ monolayer [21,22], can also be utilized to realize quantum enhanced JJFETs.

In summary, we have fabricated a Josephson junction field-effect transistor (JJFET) in a zero-energy-gap InAs/GaSb heterostructure where the excitonic insulator transition was observed. In this JJFET the superconducting critical current jumps sharply from zero to a finite value as the gate bias is varied. A gain factor of ~ 0.06 is measured, which is more than 50 times that recently reported in a conventional JJFET made of InAs quantum wells. Further optimization may allow achieving a gain factor larger than 1, needed for logic applications.

The work is supported by the LDRD program at Sandia National Laboratories, and was performed, in part, at the Center for Integrated Nanotechnologies, a U.S. Department of Energy, Office of Basic Energy Science, user facility. W.P. thanks E. Rossi for helpful discussions and M.C. Pan for useful suggestions in editing the manuscript. Sandia National Laboratories is a multimission laboratory managed and operated by National Technology and Engineering Solutions of Sandia LLC, a wholly owned subsidiary of Honeywell International Inc. for the U.S. DOE's National Nuclear Security Administration under contract DE-NA0003525. This paper describes objective technical results and analysis. Any subjective views or opinions that might be expressed in the paper do not necessarily represent the views of the U.S. DOE or the United States Government.

Figures and Figure Captions:

Fig 1

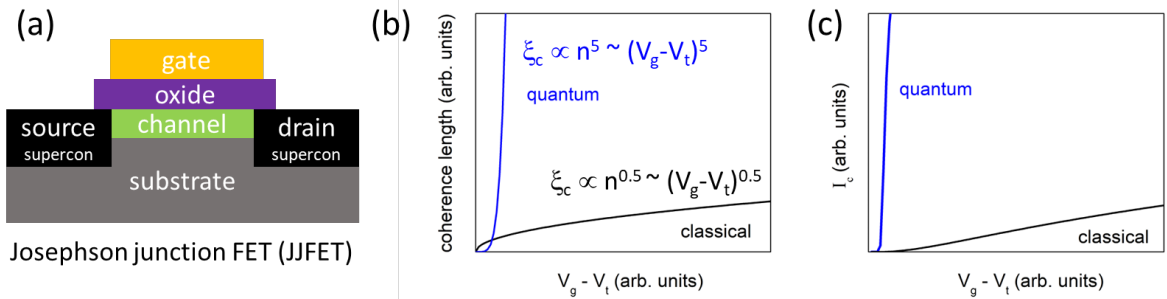

Fig. 1: (a) Schematic of a Josephson junction field-effect transistor (JJFET). The source and drain electrodes are made of superconducting material, such as tantalum. (b) Schematic gate bias dependence of coherence length in a conventional classical JJFET and in a quantum enhanced JJFET. (c) Schematic gate bias dependence of superconducting critical current for classical and quantum JJFETs.

Fig. 2

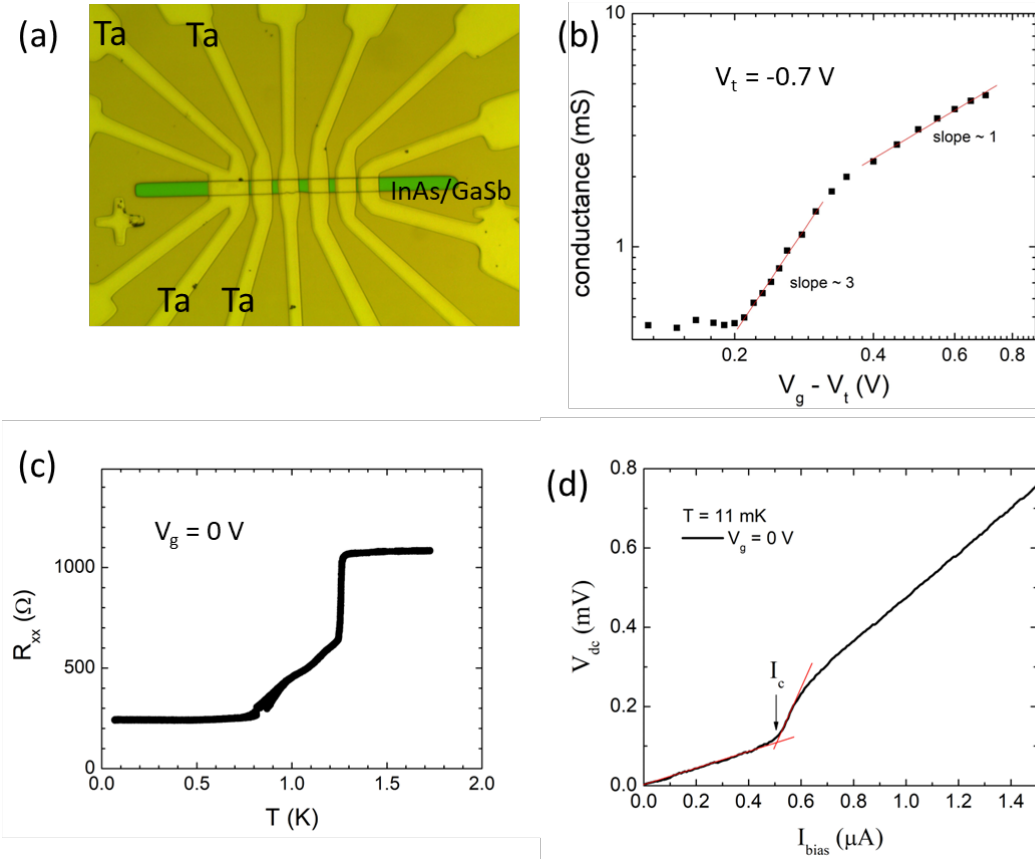

Fig. 2: (a) Optical image of a JJFET device. The green-colored bar represents the InAs/GaSb heterostructure mesas. Superconducting Ta electrodes are marked. (b) Conductance of the Josephson junction as a function of gate bias. The threshold voltage $V_t = -0.7$ V is determined from the position of resistance maximum. (c) Josephson junction resistance $R_{xx}$ at $V_g = 0$ V as a function of temperature. The sharp drop at T ~ 1.2 K is due to the onset of the superconducting transition in Ta. $R_{xx}$ continues to drop as T is lowered and saturates at a finite value of ~ 250 Ω when T < 0.8 K. (d) Current-voltage (I-V) characteristics of the Josephson junction at $V_g = 0$ V. Three sections are clearly seen. At large bias current, the I-V curve is linear. Between ~ 0.5 and 0.6 μA, it then experiences a sharp drop. Below ~ 0.5 μA, the I-V curve is linear again, but with a smaller slope (lower resistance) compared to the large bias current regime, indicating a superconducting proximity effect. A critical current $I_c$ ~ 0.5 μA, as indicated by the intersection of the two red lines in the figure, can be deduced.

Fig. 3

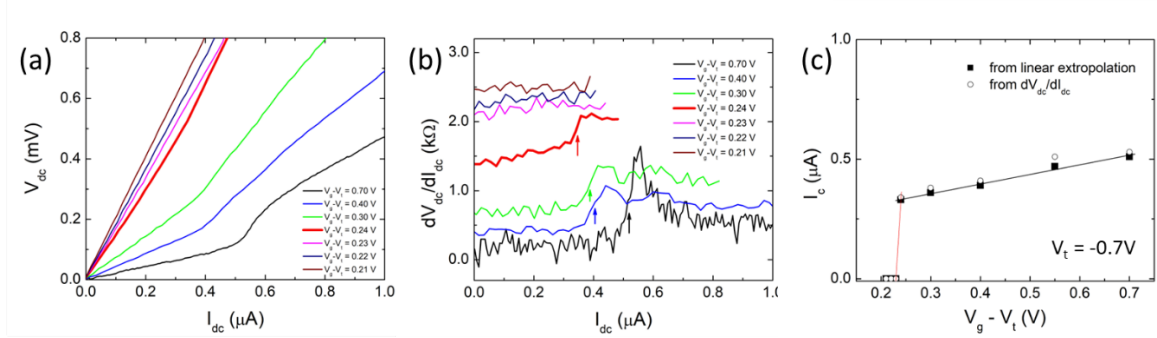

Fig. 3: (a) I-V curves at different gate bias. Above $V_g$ = -0.46 V (or $V_g$-$V_t$ = 0.24 V), a knee-like behavior is observed, from which $I_c$ can be deduced. Below -0.46 V, the I-V curve becomes linear, and the critical current is zero. (b) $dV_{dc}/dI_{dc}$ vs $I_{dc}$ at different gate bias, another method to deduce $I_c$. The arrows mark the value of $I_c$ for different $V_g$-$V_t$. The traves at $V_g$-$V_t$ = 0.21, 0.22, 0.23 V are shifted up by 0.5 kΩ, respectively, for clarity. (c) Critical current $I_c$ versus $V_g$-$V_t$, based on two different methods in (a) (solid squares) and (b) (open circles), respectively. The black and red lines are linear fits. For the red line, $dI_c/dV_g$ = 33 µA/V is obtained.